\documentclass[letterpaper]{article}
\usepackage{verbatim,graphicx,latexsym,newalg,enumerate}
\usepackage{algorithm}
\usepackage{algpseudocode}
\begin{document}
\title{Detection and Filtering of Collaborative Malicious Users in Reputation System using Quality Repository Approach}
\author{Jnanamurthy HK and Sanjay~Singh\thanks {Sanjay Singh is with the Department of Information and Communication Technology, Manipal Institute of Technology, Manipal University, Manipal-576104, INDIA \hspace{5cm}E-mail: sanjay.singh@manipal.edu}}
\maketitle

\begin{abstract}
Online reputation system is gaining popularity as it helps a user to be sure about the quality of a product/service he wants to buy. Nonetheless online reputation system is not immune from attack. Dealing with malicious ratings in reputation systems has been recognized as an important but difficult task. This problem is challenging when the number of true user's ratings is relatively small and unfair ratings plays majority in rated values. In this paper, we have proposed a new method to find malicious users in online reputation systems using Quality Repository Approach (QRA). We mainly concentrated on anomaly detection in both rating values and the malicious users. QRA is very efficient to detect malicious user ratings and aggregate true ratings. The proposed reputation system has been  evaluated through simulations and it is concluded that the QRA based system significantly reduces the impact of unfair ratings and improve trust on reputation score with lower false positive as compared to other method used for the purpose.
\end{abstract}

\section{Introduction}
Internet offers vast new opportunities to interact with strangers. These interactions can be useful, informative, even profitable. But they also involve risk. Does the product at flipkart.com \cite{fli} and Amazon.com \cite{ama} have high quality as they claim? To solve these problems, one of the most ancient mechanisms in the history of human society, word of mouth, is gaining new significance in cyberspace, where it is called reputation system \cite{two}. The online reputation systems, also called as the online feedback-based rating systems, are creating large scale, virtual word-of-mouth networks in which individuals share opinions and information by providing ratings to products, companies, hotels, digital contents etc. For example, Citysearch.com \cite{city} solicits and displays user ratings on restaurants, bars, hotels and performances. Amazon.com and YouTube.com \cite{you} recommends  products and video clips based on user's ratings.
\par
There are different defense schemes to detect malicious users and analyzing rating values to protect feedback-based reputation systems from several angles.
In Bayesian reputation systems \cite{three}\cite{four}, the reputation scores are in fact the prediction of the objects future behaviors based on the data describing their past behaviors and majority rule. An updated reputation score (i.e., posteriori) is computed based on the previous reputation score (i.e., priori) and the new feedback. There may be chance of missing to detect malicious users if malicious user rated similar to true users ratings. 
\par
Zhang and Cohen \cite{five} has proposed agents based detection of malicious users. When a consumer agent calculates the reputation score of a provider agent (users) based on advice provided by advisor agents, the advisor agents have different opinions about the provider agents reputation. The decision making is based on advisor agents opinion, so its not a proper idea to find malicious users. Laureti et al \cite{six} has estimated quality of each product, which is calculated as average of all ratings provided by users. Initially all of the ratings given to an product are assigned with equal weights and user's judging power is calculated. A user having rating values nearer to product's estimated quality is considered as a user with larger judging power. This approach is may not be feasible when ratings provided to the product is far from majority ratings. 
\par 
TAUCA \cite{ten}, a method that finds malicious users and recovers reputation scores with the help of temporal analysis and user correlation analysis. TAUCA identifies the products under attack, the time when attacks occur, and malicious users who insert dishonest ratings. TAUCA used CUSUM which is sensitive to small changes only, but not suitable for the larger changes. Feng et al \cite{eleven}, has proposed RepHi, in which attackers disguise as routers to attack and modify ratings provided by trusted users. This attack can cause undermining the reputation system scores hence manipulating the reputation. RepHi is applicable only when the participants are not familiar with each other and ratings are unencrypted. 
\par
Collaborative unfair raters through signal modeling and trust deals set of methods jointly detect collaborative unfair ratings based on signal modeling \cite{twelve}. Based on the detection, a trust-assisted rating aggregation system is designed and a trust relationship is established between two parties for a specific action. In this method, calculating reputation score is based on trusted value of each user. However it is difficult to judge exact trusted value of all users to find reputation score. 
\par
These methodologies fails to detect malicious users at early stage of threat analysis, fails to restrict malicious users who do not know about product features and restrict true users who do not know about product.

\par 
In this paper, we have proposed a new method to detect malicious users in reputation systems using Quality Repository Approach (QRA). We mainly concentrated on anomaly in both rating-values domain and the malicious users domain. In complex collusion attack, malicious users work together to reduce the reputation score by providing dishonest ratings. QRA is very efficient to detect malicious users rating and provides aggregate trustful rating.
\par
Rest of the paper is organized as follows. Section II describes the proposed method. Section III discusses the results obtained and finally section IV concludes this paper.

\section{Proposed Method}
The proposed scheme consists of following modules:
\begin{itemize}
\item Change Detector
\item Quality Repository
\item Behavior analysis
\item Aggregation algorithm
\end{itemize}

Figure.1 illustrate, the overall design of the proposed QRA's components.

\begin{figure}[bpht!]
	\centering
		\includegraphics[scale=0.4]{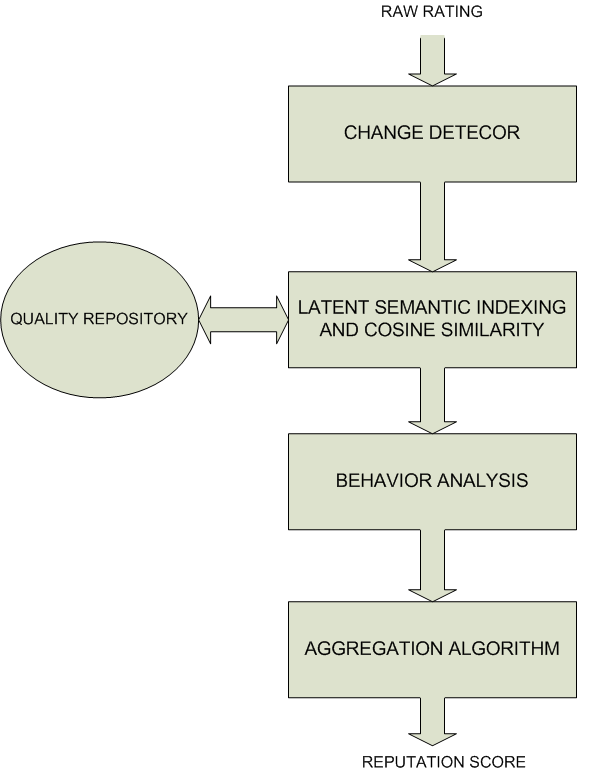}
	\caption{QRA components}
	\label{notFeather}
\end{figure}

\subsection{Change Detector}
In general, threshold is a region in which we mark a boundary for a new state. Threshold selection plays important role in finding malicious users and decision making whether user is a true user. The  selection of threshold for newly launched product is not an easy task, because we cannot predict the newly launched product whether it is a good product or a bad product.

\par 
To overcome this problem, we assume that threshold is the mean value of all possible rating values. Let X$_1$, X$_2$, X$_3$, $X_4$ and $X_5$ be the possible rating values, than
\begin{displaymath}
Initial\_Threshold =\frac{X_{1}+X_{2}+X_{3}+X_{4}+X_{5}}{5}
\end{displaymath}
 For example, in Amazon.com the possible values for rating are 1, 2, 3, 4 and 5 (5 point rating system) and the mean value is 3. If the provided rating does not match with the threshold then, analyze the behavior of the user who rated with other products, otherwise truthful user. The initial threshold value represents the quality of the product is between low and high. False alarm rate of initially launched product is more and it is effective in change detection because we cannot predict the quality of product initially.
\par     
Many statistical defense schemes consider each product equally in terms of setting detection parameters. In this paper we propose mean bisector analysis as change detector to detect malicious users.

\subsubsection{Mean Bisector Analysis} 
Selection of mid-value is the point in which the distance from lowest rated value and maximum rated value to that point is almost same. The mid-value plays important role in selection of threshold and analyze how far the rated values present from mid-point. Procedure to find threshold selection and rated data analysis is given below.

\begin{enumerate}[(i)]
\item Initially find mean value of all ratings provided by the users, where low is zero, high is the number of raters rated to particular product and $r{_i}$ is the $i^{th}$ rated value.
\begin{center}
 $\mu$=$\displaystyle\frac{1}{{high-low}}$  $\displaystyle\sum_{i=low}^{high} r{_i} $
\end{center}
\item Bisect the rating values into two groups i.e Group1 and Group2 ( R$_1$ and R$_2$ ) using mean $\mu$  where R$_1$$<$$\mu$ \& R$_2$$\geq$$\mu$.
\item   Calculate the mean values of Group1, Group2 i.e  $\mu$$_1$   and  $\mu$$_2$ 
\begin{center}
	$\mu$$_1$=$\displaystyle\frac{1}{{{\mu}-low}}$  $\displaystyle\sum_{i=low}^{\mu} r{_i} $ \\
	$\mu$$_2$=$\displaystyle\frac{1}{{high-{\mu}}}$  $\displaystyle\sum_{i={\mu}+1}^{high} r{_i} $
\end{center}
\item  Find the  Mid-value, from $\mu$$_1$ and $\mu$$_2$ calculated from above equations.
\begin{displaymath}
         Mid-value=\frac{{\mu{_1}+\mu{_2}}}{2}
\end{displaymath}
\item Repeat steps (ii)-(iv) until the Mid-value  or $\mu$$_1$ and $\mu$$_2$ in successive iterations do not  change.

\item Threshold selection is significant in change detection and malicious user detection. Here we select threshold values based on statistical methods i.e how the
rated data deviated from mean value.  
\begin{displaymath}
deviation=\sqrt{\frac{\sum\limits_{i=1}^{n} \left(r_{i} - \bar{r}\right)^{2}}{n}} + {sensitivity}
\end{displaymath}

where sensitivity factor represents how strongly a system detects the changes in the rated data. When the value of sensitivity factor is small then it is more sensitive to change detection whereas it is less sensitive to change detection for the large value of sensitivity factor. We have checked $-0.4$, $-0.2$, 0  and $+0.2$ as sensitivity factor. The value between $-0.4$ and $+0.2$ is effective in change detection.

\item Thresholds: Upgrading threshold is a margin for change detection of self boosting of reputation scores and downgrading threshold is a margin for change detection in downgrading of reputation scores.
\end{enumerate}

$\mbox{Upgrading Threshold} =\mbox{Mid-value} + deviation$\\
$\mbox{Downgrading Threshold}=\mbox{Mid-value} - deviation$

\subsubsection{Alarm}
Point at which currently rated value exceeds upgrading or downgrading threshold.\\
\begin{displaymath}
 t_a=min\{ k| r_k^+ \geq \mbox{Upgrading Threshold}\}
\end{displaymath}

\begin{displaymath}
 t_a=min\{k| r_k^-\leq \mbox{Downgrading Threshold}\}
\end{displaymath}
where
\begin{itemize}
\item $r_k^+$ is the rating value of  $k_{th}$  data sample ($+$ represents upgrading).
\item $r_k^-$ is the rating value of  $k_{th}$  data sample ($-$ represents downgrading).
\item $t_a$ is  the point when the detector identifies a change and raises an alarm.
\end{itemize}

\begin{figure}[bpht!]
	\centering
		\includegraphics[height=250px]{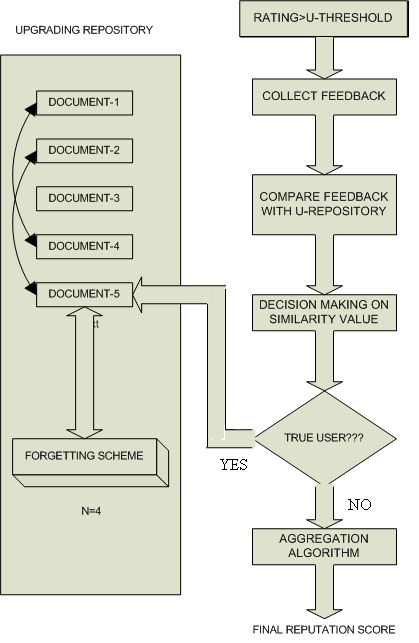}
	\caption{Upgrading-Repository}
	\label{Feather}
\end{figure}

\subsection{Quality Repository}
Product quality repository represents quality of product, that collects the feedbacks provided by the true users.  Latent Semantic Indexing (LSI) \cite{lsi} and Cosine similarity \cite{Raghavan} has been used to create Quality Repository dynamically based on feedback provided by the true users.\\ 
Quality Repository is classified into two categories
\begin{itemize}
\item \textit{Upgrading-Repository} collects feedbacks provided by the true users who boost the reputation scores (Positive opinion about products). Collect the feedback if user rating value exceeds Upgrading-threshold, and find the similarity with product Upgrading-Repository. Decision making about users based on similarity value is shown in Fig.2. Similarity value of feedback and Quality repository plays important role to group the users. Selection of similarity threshold is based on false alarm rate.
\item \textit{Downgrading-Repository} collects feedbacks provided by the true users who downgrade the reputation scores. Collect the feedback if user rating value as lower Downgrading-threshold, and find the similarity with product Downgrading-Repository. Decision making about users based on similarity value is shown in Fig.3. Similarity value of feedback and Quality repository plays important role to group the users.
\end{itemize}
It is to be noted that initial Quality Repository is the quality information given by the product manufacturer.

\begin{figure}[bpht!]
	\centering
		\includegraphics[height=250px]{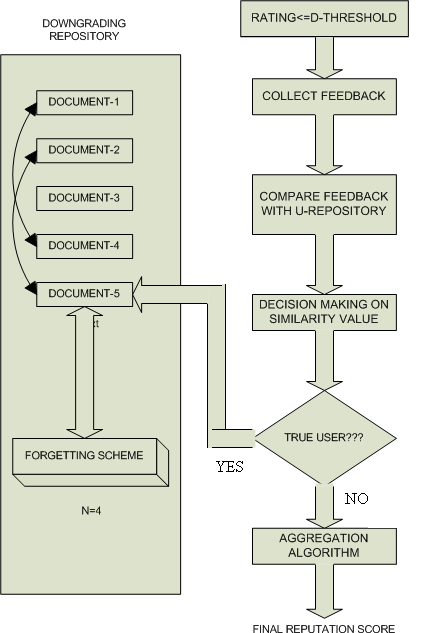}
	\caption{Downgrading-Repository}
	\label{Feather}
\end{figure}

\subsubsection{Indexing Feedbacks in Repository}
Latent Semantic Indexing (LSI) has been used to compute the similarity score between the documents in quality repository and the feedbacks provided by the users. Term-document matrix for the feedbacks is created in quality repository and query matrix for the feedback provided by the suspected user. Following steps are followed to index the user feedback in repository:
\begin{itemize}
\item Decompose the term document matrix of quality repository(Q$_{rep}$)into three new matrices \textbf{U, S, V} and then find \textbf{V$^{T}$}
\begin{center}
\textbf{Q$_{rep}$=USV$^{T}$}
\end{center}
\item Dimensionality Reduction: Computing U$_{k}$, S$_{k}$, V$_{k}$ and V$_{k}$$^{T}$, where k is rank of approximation.
\begin{center}
\textbf{Q$_{rep}$$^{T}$=(USV$^{T}$)$^{T}$= VSU$^{T}$}
\end{center}

\begin{center}
\textbf{Q$_{rep}$$^{T}$US$^{-1}$= VSU$^{T}$US$^{-1}$}
\end{center}

\begin{center}
\textbf{V=Q$_{rep}$$^{T}$US$^{-1}$}
\end{center}

\item V consists of n rows, each row containing the coordinates of a document vector. For a given document vector Q$^{rep}$ above equation can be rewritten as
\begin{center}
\textbf{Q$^{rep}$=Q$_{rep}$$^{T}$US$^{-1}$}
\end{center}

\item Since in LSI a query q is treated just as another document then the query vector as feedback provided by the user is given by
\begin{center}
\textbf{feedback=q$^{T}$US$^{-1}$}
\end{center}
\item Reduced k-dimensional space we can write
\begin{center}
\textbf{Q$^{rep}$=Q$_{rep}$$^{T}$U$_{k}$S$_{k}$$^{-1}$}
\end{center}
\begin{center}
\textbf{feedback=q$^{T}$U$_{k}$S$_{k}$$^{-1}$}
\end{center}

\item Query-document similarity is computed using cosine similarity\\
\textbf{sim(feedback, Q$^{rep}$)=sim(q$^{T}$U$_{k}$S$_{k}$$^{-1}$,Q$_{rep}$$^{T}$U$_{k}$S$_{k}$$^{-1}$)} 
\end{itemize}
Based on similarity value of feedbacks, users are grouped into 3 levels (Level 1, Level 2 and Level 3). Selection of similarity threshold depends on false alarm rate. 

\subsubsection{Forgetting Scheme}
Old feedback provided by the users may not always be relevant for reputation rating, because the product launcher may change
quality of product over time. Old feedbacks are not considered to represent quality of the product, this translates into gradually forgetting old feedbacks in quality repository. This can be achieved by introducing a forgetting factor, selection of forgetting factor is based on rapidity of change in product quality.

\subsection{Behavior Analysis}
In QRA (proposed method) we are using users behavior analysis to find malicious users. Behavior analysis is a technique that can show whether and how strongly one user is similar with other users. 

\begin{enumerate}
\item \textit{Behavior analysis of single user with other products} (Assuming all the products are of different quality): When the user provides ratings to a particular product, data analysis module analyze the current rated value with genuine users ratings, if there is a difference in the currently rated value, behavior analysis module checks how the user is  behaving with other products. 
\item \textit{Behavior analysis of multiple user IDs with commonly rated products}: When the user provides ratings to a particular product, data analysis module analyze the currently rated value with genuine users ratings, if there is a difference in the currently rated value, behavior analysis module checks how the current user IDs ratings is similar with other user IDs.  
\end{enumerate}
To analyze behavior of users A and B, we used cosine similarity method.  A and B is the rating values of common user IDs rated to common products, the cosine similarity, $\theta$ is represented as

\begin{displaymath}
         \mbox{similarity} = cos(\theta)=\frac{\vec{V}(A).\vec{V}(B)}{|\vec{V}(A)|.|\vec{V}(B)|}
\end{displaymath}

\section{Results and Discussion}
We assumed few user IDs as malicious users and conducted experiment on heterogeneous products. In QRA the reputation score is calculated dynamically to newly launched products and it is found that the proposed method is very efficient to detect the malicious users.  
\par
Malicious users aim is to reduce and manipulate reputation score of other products and true users aim is to provide honest ratings to the products. Reputation score calculated based on honest users rating values indicates the quality of the product.

\begin{table*}[t]
\centering
\caption{Users status and Reputation scores}
\label{t1}
\begin{tabular}{|c|c|c|c|c|c|}
\hline
UID&Rating&Self Boosting Threshold&Downgrading Threshold& Final Rating& User Status\\ \hline
1&3&3.000&3.000&3.00& T-USER \\ \hline 
2&5&3.000&3.000&4.00&T-USER \\ \hline
3&1&5.414&2.585&4.00&\textbf{M-USER} \\ \hline
4&5&5.414&2.585&4.33&T-USER \\ \hline
5&2&5.154&2.845&4.33&\textbf{M-USER} \\ \hline
6&3&5.154&2.845&4.00&T-USER \\ \hline
7&4&5.154&2.845&4.00&T-USER \\ \hline
8&5&4.833&2.833&4.16&T-USER \\ \hline
9&4&4.858&2.891&4.14&T-USER \\ \hline
10&1&4.699&2.900&4.14&\textbf{M-USER} \\ \hline
11&4&4.699&2.900&4.12&T-USER \\ \hline
12&5&4.584&2.915&4.22&T-USER \\ \hline
13&5&5.133&3.466&4.30&T-USER \\ \hline
14&1&5.123&3.476&4.30&\textbf{M-USER} \\ \hline
15&5&5.123&3.476&4.36&T-USER \\ \hline
\end{tabular}
\end{table*}

\par In QRA we have used mean bisector analysis as change detector. Mean bisector analysis plays very important role in the selection of mid-value, threshold and to analyze how far the rated values are from mid-point. Table.I shows UID, corresponding rating provided by UID, upgrading threshold, downgrading threshold, final rating (reputation score) and status of the user whether the user is malicious or not. 
\par 
In row one of Table.I shows UID 1 rated 3 to a particular product, the rated value 3 is similar to the downgrading and upgrading threshold, then it is concluded as normal rating and status of the user is true. Row two shows UID 2 rated 5 to a particular product, the rated value 5 exceeds the upgrading threshold, then  quality repository module concluded it as normal rating, and status of the user is true. Row three of shows UID 3 rated 1 to a particular product, the rated value 1 which is less than the downgrading threshold, then quality repository module concluded it as malicious rating.  Status of the user is malicious and final rating (reputation score) remains unchanged.
\par
User IDs 3, 5, 10 and 14 are users whose rated values are lower than the downgrading thresholds. On further asking for textual feedbacks and comparison with quality repository concluded these as malicious users.


Fig.4 shows points at which rated values exceeds the threshold value. In position 3, 5, 10 and 14 rated values lower than the downgrading threshold (0.0 considered as sensitivity factor). Further verification based on textual feedback is required to ascertain that these users are malicious users and responsible for downgrading the reputation score. This inference is confirmed by comparing the textual feedback with quality repository, and decision is made on similarity score. In QRA analysis the rated value at the position 3, 5, 10 and 14 are concluded as malicious users. 

\begin{figure}[bpht!]
	\centering
		\includegraphics[width=8cm]{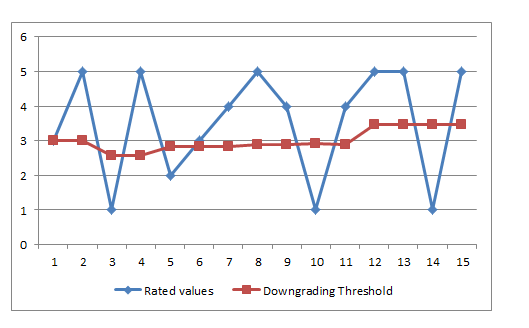}
		\caption{Rated data v/s Thresholds}
	\label{fig:Reputation system structure}
\end{figure}

\begin{figure}[bpht!]
	\centering
		\includegraphics[width=8cm]{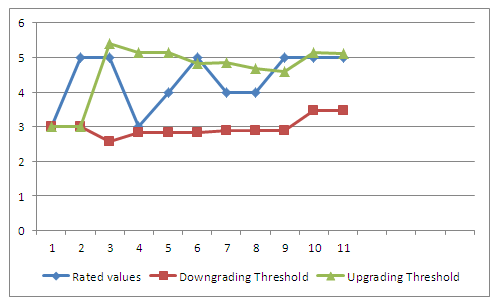}
		\caption{Rated data v/s Thresholds}
	\label{fig:Reputation system structure}
\end{figure}

\begin{table}
\centering
\caption{Upgrading-Repository}
\begin{tabular}{||c||c||}
\hline
\hline
doc 1 & "battery backup yields more than 2 hrs, so it is very good" \\  \hline\hline
doc 2 & "touch screen is very good and display also good" \\  \hline\hline
doc 3 & "battery backup yields more so it is very good" \\  \hline\hline
doc 4 & "touch screen is very good and display good" \\  \hline\hline
\end{tabular}
\end{table}

\begin{table}
\centering
\caption{Coordinate values of Feedback and Quality Repository using SVD}
\begin{tabular}{||c||c||c||c||c||}
\hline
\hline
Feedback & doc 1 & doc 2 & doc 3 & doc 4 \\  \hline\hline
-0.3967 & -0.5123 & -0.5231 & -0.4576 & -0.5045 \\  \hline\hline
0.3877  & -0.5816 &  0.5053 & -0.4375 &  0.4636\\  \hline\hline
\end{tabular}
\end{table}

\begin{table}
\centering
\caption{Document and Similarity value }
\begin{tabular}{||c||c||}
\hline
\hline
Document-ID & Similarity \\  \hline\hline
2  &  1.0000 \\  \hline\hline
4  &  0.9995 \\  \hline\hline
3  &  0.0338 \\  \hline\hline
1  & -0.0518 \\  \hline\hline
\end{tabular}
\end{table}

Figure.5 shows points at which rated values exceeds the threshold value. In position 2, 6 and 9 rated values exceeds upgrading threshold, further verification is required, which is based on textual feedback. Textual feedback is compared with the quality repository of the product, and decision is made based on the similarity score.  For the given example at the position 2, 6 and 9 rated value exceeds upgrading threshold, then reputation system requesting user to provide feedback for boosting the reputation score. For given example, user feedback "touch screen and display is very good"  given by user is compared with the upgrading quality repository. Upgrading quality repository contains feedbacks provided by true users given and shown in Table II.
\par
The vector component obtained after LSI and feedback from the user is given in Table III. Cosine similarity between the feedback provided by user and quality repository is given in Table IV. Document 2 yields similarity value 1 and document 4 yields 0.999, i.e feedback provided by the user is almost similar to that particular document. Then we conclude that user is a true user. Total aggregation rating excluding malicious users is 4.3600 and total aggregation rating including malicious users is 3.5333. The difference between including malicious users and excluding malicious users rating is 0.8267 (positive value), this represents attack type is downgrading.
\par 
To demonstrate the overall performance of QRA, we compared the proposed scheme with anomaly detection in feedback based reputation systems through temporal and correlation analysis \cite{ten}, a method that finds change detection based on CUSUM method. CUSUM is sensitive to small changes in rating values, this treats true users also as malicious users and false alarm rate is more. We conducted experiment for 10, 15, 20 and 25 raters using both QRA and CUSUM method for the same set of data, the proposed method causes less false alarm rate than the CUSUM change detection and the result is shown in Table.V. QRA method is very efficient to detect collaborative malicious users who insert unfair ratings for upgrading or downgrading the product rating values in reputation system.

\begin{table}
\centering
\caption{False alarm rate of Proposed Method versus CUSUM Method }
\begin{tabular}{||c||c||c||}
\hline
\hline
Number of Users & Proposed Method & CUSUM Method\\  \hline\hline
0   &  0.00 & 0.00\\  \hline\hline
10  &  0.33 & 0.66\\  \hline\hline
15  &  0.25 & 0.50\\  \hline\hline
20  &  0.37 & 0.40\\  \hline\hline
25  &  0.37 & 0.45\\  \hline\hline
\end{tabular}
\end{table}

\section{Conclusion}
Importance of online reputation system has increased in recent time. Online reputation system gives clue about the quality of a product or service. However there is a chance of attack on reputation system to either degrade the reputation score or boost the reputation score for a particular product/service.
In this paper, we have proposed QRA based method which is very efficient to detect collaborative malicious users who insert unfair ratings for upgrading or downgrading the product rating values in reputation systems. The basic idea is to integrate the anomaly detection based on heterogeneous thresholds and analyze the feedbacks of users. This method uses statistical change detection of ratings, similarity among feedbacks calculated using Latent Semantic Indexing (LSI) and cosine similarity which gives the result varies which varies between -1 and 1. The experiments showed that the QRA system accurately detects colluded malicious raters and therefore significantly improves the trustfulness of reputation systems.

\bibliographystyle{IEEEtran}
\bibliography{myref}

\begin{thebibliography}{10}
\providecommand{\url}[1]{#1}
\csname url@samestyle\endcsname
\providecommand{\newblock}{\relax}
\providecommand{\bibinfo}[2]{#2}
\providecommand{\BIBentrySTDinterwordspacing}{\spaceskip=0pt\relax}
\providecommand{\BIBentryALTinterwordstretchfactor}{4}
\providecommand{\BIBentryALTinterwordspacing}{\spaceskip=\fontdimen2\font plus
\BIBentryALTinterwordstretchfactor\fontdimen3\font minus
  \fontdimen4\font\relax}
\providecommand{\BIBforeignlanguage}[2]{{%
\expandafter\ifx\csname l@#1\endcsname\relax
\typeout{** WARNING: IEEEtran.bst: No hyphenation pattern has been}%
\typeout{** loaded for the language `#1'. Using the pattern for}%
\typeout{** the default language instead.}%
\else
\language=\csname l@#1\endcsname
\fi
#2}}
\providecommand{\BIBdecl}{\relax}
\BIBdecl

\bibitem{fli}
, ``Flipkart.com-is an {I}ndian e-commerce company,'' [Online] Available:
  http://www.flipkart.com/.

\bibitem{ama}
, ``Amazon.com-is an {A}merican multinational e-commerce company,'' [Online]
  Available: http://www.amazon.com/.

\bibitem{two}
C.~Dellarocas, ``The digitization of word-of-mouth: Promise and challenges of
  online reputation systems,'' 2001.

\bibitem{city}
O.~Available:, ``www.citysearch.com.''

\bibitem{you}
C.~Hurley, S.~Chen, and J.~Karim, ``Youtube is a video sharing website,''
  [Online] Available: www.youtube.com.

\bibitem{three}
L.~Mui, M.~Mohtashemi, C.~Ang, P.~Szolovits, and A.~Halberstadt, ``Ratings in
  distributed systems: A bayesian approach,'' 2001.

\bibitem{four}
B.~E. Commerce, A.~Jøsang, and R.~Ismail, ``The beta reputation system,'' in
  \emph{In Proceedings of the 15th Bled Electronic Commerce Conference}, 2002.

\bibitem{five}
J.~Zhang and R.~Cohen, ``A personalized approach to address unfair ratings in
  multiagent reputation systems,'' in \emph{Europhysics Letters 75(6):
  1006-1012, 2006}, 2006.

\bibitem{six}
P.~Laureti, L.~Moret, Y.~C. Zhang, and Y.-K. Yu, ``Information filtering via
  iterative refinement,'' in \emph{in Proc. of the Fifth Int. Joint Conf. on
  Autonomous Agents and Multiagent Systems (AAMAS) Workshop on Trust in Agent
  Societies}, 2006.

\bibitem{ten}
\BIBentryALTinterwordspacing
Y.~Liu and Y.~L. Sun, ``Anomaly detection in feedback-based reputation systems
  through temporal and correlation analysis,'' in \emph{Proceedings of the 2010
  IEEE Second International Conference on Social Computing}, ser. SOCIALCOM
  '10.\hskip 1em plus 0.5em minus 0.4em\relax Washington, DC, USA: IEEE
  Computer Society, 2010, pp. 65--72. [Online]. Available:
  \url{http://dx.doi.org/10.1109/SocialCom.2010.19}
\BIBentrySTDinterwordspacing

\bibitem{eleven}
J.~Feng, Y.~Zhang, S.~Chen, and A.~Fu, ``Rephi: A novel attack against p2p
  reputation systems,'' in \emph{Computer Communications Workshops (INFOCOM
  WKSHPS), 2011 IEEE Conference on}, april 2011, pp. 1088 --1092.

\bibitem{twelve}
\BIBentryALTinterwordspacing
Y.~Yang, Y.~L. Sun, S.~Kay, and Q.~Yang, ``Defending online reputation systems
  against collaborative unfair raters through signal modeling and trust,'' in
  \emph{Proceedings of the 2009 ACM symposium on Applied Computing}, ser. SAC
  '09.\hskip 1em plus 0.5em minus 0.4em\relax New York, NY, USA: ACM, 2009, pp.
  1308--1315. [Online]. Available:
  \url{http://doi.acm.org/10.1145/1529282.1529575}
\BIBentrySTDinterwordspacing

\bibitem{lsi}
S.~Deerwester, S.~T. Dumais, G.~W. Furnas, T.~K. Landauer, and R.~Harshman,
  ``Indexing by latent semantic analysis,'' \emph{Journal of The American
  Society For Information Science}, vol.~41, no.~6, pp. 391--407, 1990.

\bibitem{Raghavan}
C.~D. Manning, P.~Raghavan, and H.~Sch\"{u}tze, \emph{Introduction to
  Information Retrieval}.\hskip 1em plus 0.5em minus 0.4em\relax New York, NY,
  USA: Cambridge University Press, 2008.

\end{thebibliography}
\end{document}